\documentclass[aps,prc,twocolumn,superscriptaddress,showpacs]{revtex4-1}

\usepackage{graphicx}
\usepackage{dcolumn}
\usepackage{bm}
\usepackage{ulem}
\usepackage{color}
\usepackage{epsfig}

\newcommand{\xeag}{$^{124}$Xe($\alpha$,$\gamma$)$^{128}$Ba}
\newcommand{\xean}{$^{124}$Xe($\alpha$,n)$^{127}$Ba}
\newcommand{\xeap}{$^{124}$Xe($\alpha$,p)$^{127}$Cs}
\newcommand{\rag}{($\alpha$,$\gamma$)}
\newcommand{\ran}{($\alpha$,n)}
\newcommand{\rap}{($\alpha$,p)}
\newcommand{\rga}{($\gamma$,$\alpha$)}
\newcommand{\rgp}{($\gamma$,p)}
\newcommand{\rgn}{($\gamma$,n)}
\newcommand{\rng}{(n,$\gamma$)}
\newcommand{\rpg}{(p,$\gamma$)}

\begin{document}

\title{
  Experimental study of the astrophysical $\gamma$-process reaction $^{124}$Xe($\alpha,\gamma$)$^{128}$Ba
}

\author{Z. Hal\'asz}%
\affiliation{Institute for Nuclear Research (Atomki), H-4001 Debrecen, Hungary}
\affiliation{University of Debrecen, H-4001 Debrecen, Hungary}
\author{E. Somorjai}%
\affiliation{Institute for Nuclear Research (Atomki), H-4001 Debrecen, Hungary}
\author{Gy. Gy\"urky}%
\email{gyurky@atomki.mta.hu}
\affiliation{Institute for Nuclear Research (Atomki), H-4001 Debrecen, Hungary}
\author{Z. Elekes}%
\affiliation{Institute for Nuclear Research (Atomki), H-4001 Debrecen, Hungary}
\author{Zs. F\"ul\"op}%
\affiliation{Institute for Nuclear Research (Atomki), H-4001 Debrecen, Hungary}
\author{T. Sz\"ucs}
\affiliation{Institute for Nuclear Research (Atomki), H-4001 Debrecen, Hungary}
\author{G.G. Kiss}%
\affiliation{Institute for Nuclear Research (Atomki), H-4001 Debrecen, Hungary}
\author{N. Szegedi}%
\affiliation{University of Debrecen, H-4001 Debrecen, Hungary}
\author{T. Rauscher}%
\affiliation{Centre for Astrophysics Research, School of Physics, Astronomy and Mathematics,
University of Hertfordshire, Hatfield AL10 9AB, United Kingdom}
\affiliation{Department of Physics, University of Basel, 4056 Basel, Switzerland}
\author{J. G\"orres}%
\affiliation{Nuclear Science Laboratory, University of Notre Dame, Notre Dame, Indiana, USA}
\author{M. Wiescher}%
\affiliation{Nuclear Science Laboratory, University of Notre Dame, Notre Dame, Indiana, USA}

\date{\today}

\begin{abstract}
\begin{description}

\item[Background]

The synthesis of heavy, proton rich isotopes in the astrophysical $\gamma$ process proceeds through photodisintegration reactions. For the improved understanding of the process, the rates of the involved nuclear reactions must be known. The reaction $^{128}$Ba($\gamma$,$\alpha$)$^{124}$Xe was found to affect the abundance of the $p$ nucleus $^{124}$Xe in previous rate variation studies. 

\item[Purpose]

Since the stellar rate for this reaction cannot be determined by a measurement directly, the aim of the present work was to measure the cross section of the inverse $^{124}$Xe($\alpha$,$\gamma$)$^{128}$Ba reaction and to compare the results with statistical model predictions used in astrophysical networks. Modified nuclear input can then be used to provide an improved stellar reaction rate. Of great importance is the fact that data below the \ran\ threshold was obtained. Studying simultaneously the $^{124}$Xe($\alpha$,n)$^{127}$Ba reaction channel at higher energy allowed to further identify the source of a discrepancy between data and prediction.

\item[Method]

The $^{124}$Xe($\alpha$,$\gamma$)$^{128}$Ba and $^{124}$Xe($\alpha$,n)$^{127}$Ba cross sections were measured with the activation method using a thin window $^{124}$Xe gas cell and an $\alpha$ beam from a cyclotron accelerator. The studied energy range was between E$_\alpha$\,=\,11 and 15 MeV close above the astrophysically relevant energy range.

\item[Results]

The obtained cross sections are compared with Hauser-Feshbach statistical model calculations. The experimental cross sections are smaller than standard predictions previously used in astrophysical calculations. As dominating source of the difference, the theoretical $\alpha$ width was identified. The experimental data suggest an $\alpha$ width lower by at least a factor of 0.125 in the astrophysically important energy range.

\item[Conclusions]

An upper limit for the $^{128}$Ba($\gamma$,$\alpha$)$^{124}$Xe stellar rate was inferred from our measurement.
The impact of this rate and lower rates was studied in two different models for core-collapse supernova explosions of 25 $M_\odot$ stars. A significant contribution to the $^{124}$Xe abundance via this reaction path would only be possible when the rate was increased above the previous standard value. Since the experimental data rule this out, they also demonstrate the closure of this production path.

\end{description}
\end{abstract}


\maketitle

\section{Introduction}
\label{sec:intro}

The chemical elements in the upper half of the periodic table represent only a tiny fraction of matter building up our universe. The reason for the rarity of these heavy elements is that they are produced in stars in subsidiary processes, not in the ones powering the star. It has been shown that these processes involve neutron-induced reactions that build the heavy elements from the Iron group up to Uranium. Two distinct processes, characterized by neutron density and duration, have been identified: the slow and the rapid neutron-capture process, also called $s$- and $r$ process, respectively \cite{ctt,thi11,kap11,arn07}. 

Some of the most neutron deficient isotopes of the heavy elements, moreover, are even less abundant than the more neutron-rich ones. These stable isotopes on the proton rich side of the valley of stability between $^{74}$Se and $^{196}$Hg are the so-called $p$ isotopes. They are not produced in any of the neutron capture processes and their isotopic abundances are typically only 0.1\,--\,1\,\% of that of their more neutron-rich neighbours. Owing to their rareness and the fact that no heavy element has a dominant $p$ isotope, their existence is identified so far only in the Solar System. We have no information on their abundance in other stars.

Regardless of their rarity, the existence of $p$ isotopes requires an explanation of their production processes and sites. To date, there is no unique and generally accepted process which could explain the abundances of \textit{all} $p$ nuclei. Several possible astrophysical production sites are suggested and also contributions from several mechanisms are considered. For reviews, see \cite{arn03,rau13} and references therein.

Perhaps the most important process which contributes to the synthesis of $p$ isotopes is the so-called $\gamma$ process which converts heavy, neutron-rich species into $p$ isotopes through photodisintegration reactions \cite{rau13}. In the reaction network of a $\gamma$-process model the \rgn\ reactions, which drive the material toward the proton rich side, play the key role. On the proton rich side, \rgp\ and \rga\ reactions start to compete with the \rgn\ reactions and influence the resulting abundances of $p$ isotopes or their $\beta$-decay progenitors. 

Although recent years have seen significant progress in modeling the $\gamma$ process in massive stars \cite{rau13,rayet95,rau02,haya08}, no good reproduction of abundances across the measured range has been achieved by the models, yet. Especially severe differences can be found between calculated and observed abundances in certain mass regions, like around the Mo-Ru isotopes or at mass numbers around 150. The failure of the models can in part be attributed to the nuclear physics input, especially in the region $A\approx 150$. For the light $p$ nuclides, thermonuclear supernovae with enhanced $s$-process seeds in binary systems have been suggested as a promising alternative \cite{howmey,kusa11,tra11,tra15}.

Lacking experimental data in the relevant mass and energy range, the rates of reactions are taken from theoretical calculations based on the Hauser-Feshbach statistical model. An experimental check of the theoretical cross sections is therefore important to examine one possible reason of the deficiencies. This need triggered huge experimental work in the last decade on studying reactions relevant for the $\gamma$ process (see the references in the review paper \cite{rau13} and some of the most recent studies \cite{har13,net13,qui13,sim13,spy13,net14,kis14,gyu14,gyu14b,yal15,gur15,kis15,net15,sim15,qui15,net15b,naq15,har16}). 

Owing to the size of a $\gamma$-process network which involves thousands of reactions, it is difficult to identify a key reaction which should be the target of experimental investigations. Sensitivity studies, however, have been carried out to point out those reactions to which $\gamma$-process models are especially sensitive, i.e., the $\gamma$-process flow and therefore the resulting abundance of certain $p$ isotopes depend strongly on the rate of these reactions.
The two, currently available sensitivity studies \cite{rau06,rap06} come up with lists of key reactions largely different from each other. This is due to the different approaches adopted. While the study by \cite{rap06} collectively varied groups of reactions to find their impact on final $p$ abundances, the paper by \cite{rau06} identified nuclides where predicted \rgn\, \rgp\, and/or \rga\ rates are close and thus a change in either rate would affect deflections of the $\gamma$-process path. There are only few reactions which are identified as being important in both studies, the $^{128}$Ba($\gamma$,$\alpha$)$^{124}$Xe reaction being among them.
Earlier experiments have shown that in the case of \rag\ reactions the discrepancy between experimental and calculated cross sections (used to obtain reaction rates) can easily be as much as a factor of three, or even an order of magnitude. Therefore, the experimental study of the identified key reaction is of high importance.

Instead of directly studying photodisintegration reactions, a cross section measurement of the reverse reaction, i.e., particle capture, is usually preferred. Technically, it is much easier to measure a charged particle capture cross section than a $\gamma$ induced one. Moreover, the effect of the thermally populated excited states of nuclei in the hot stellar plasma is much more pronounced in the case of $\gamma$-induced reactions than for captures \cite{rau13,rau12,rau13supp,rauadv}. Thus, a measurement in the direction of the capture reaction provides more relevant astrophysical information compared to the study of the $\gamma$-induced reaction itself. The aim of the present work is therefore the measurement of the \xeag\ cross section. Nevertheless, reactions on nuclei in excited states are particularly important at the high plasma temperatures encountered in the $\gamma$ process also for captures. Therefore, a measurement of laboratory ground-state (g.s.) cross sections cannot fully constrain the stellar reaction rate but a comparison to the calculated g.s.\ cross sections allows to perform an important test of the prediction.

This paper is organized as follows. Section \ref{sec:invesreac} provides some further details about the investigated reactions. The experimental technique is described in Sec.\ \ref{sec:exp}. The results are presented and discussed in Sec.\ \ref{sec:results} while conclusions and a summary are given in Sec.\ \ref{sec:conclusions}.

\section{The investigated reactions}
\label{sec:invesreac}

At typical $\gamma$-process temperatures of about 1.5 -- 3.5 GK, the relevant energy range of the \xeag\ reaction lies between 4.4\,MeV and 10.6\,MeV \cite{rau10}. Due to the high Coulomb barrier, the \xeag\ cross section is very small in this energy range, dropping from about 10 $\mu$barn down to 10$^{-17}$ barn with decreasing energy. Therefore, it was only possible to measure the cross section slightly above this energy range, between 10.6 and 13.6 MeV in center-of-mass energy. 

Besides radiative capture, several reaction channels are open in the $^{124}$Xe + $\alpha$ system in this energy range. However, as most of these channels involve the emission of low-energy charged particles, and are therefore suppressed by the Coulomb barrier, only the ($\alpha,\gamma$), ($\alpha$,n), and ($\alpha$,p) channels have non-negligible cross sections.
These three channels lead to radioactive residual nuclei, $^{128}$Ba, $^{127}$Ba and  $^{127}$Cs, respectively, and thus their cross sections can in principle be measured using the activation method. The activation method proved to be an extremely powerful technique for cross section measurements relevant for the $\gamma$ process. Above the Fe region, most of the charged-particle induced reaction cross sections have been measured with this method \cite{KADONIS}. Also in the present work, the $\alpha$-induced cross sections of $^{124}$Xe were measured with the activation method. 

\begin{table*}
\centering
\caption{\label{tab:decaypar} Decay parameters of the studied reaction products taken from the nuclear data compilation \cite{NDS128,NDS127}. Only those $\gamma$ transitions are listed which were used for the analysis.}
\begin{tabular}{llccc}
\hline
Isotope  & \multicolumn{1}{c}{Produced by} & Half-life  & E$_\gamma$ [keV] & I$_\gamma$ [\%]\\
&  & [hour] &  &  \\
\hline
$^{128}$Ba & \xeag\ & 58.32\,$\pm$\,1.2 & 273.44 & 14.5\,$\pm$\,0.7 \\
$^{127}$Ba & \xean\ & 0.212\,$\pm$\,0.007 & \multicolumn{2}{c}{not used in the analysis} \\
$^{127}$Cs & \xeap\, $^{127}$Ba($\beta^+$) & 6.25\,$\pm$\,0.1  & 411.95 & 62.9\,$\pm$\,1.0 \\
\hline
\end{tabular}
\end{table*}

The relevant decay parameters of the reaction products are summarized in Tab.\,\ref{tab:decaypar}. The measurement of the \xeap\ cross section was unfortunately hindered by the fact that $^{127}$Cs is not only populated by the ($\alpha$,p) reaction channel, but also by the $\beta$-decay of $^{127}$Ba, which is the reaction product of the \xean\ reaction. Statistical model calculations show that the \xean\ cross section is at least a factor 25 higher than that of the \xeap\ reaction from close above the \ran\ threshold. Therefore, the population of $^{127}$Cs is dominated by the $^{127}$Ba $\beta$-decay following the \xean\ reaction. Hence, it was not possible to measure the \xeap\ cross sections in the present work.

Exploiting the fact that the \xeap\ cross section is much lower than that of \xean\, the latter cross section was determined from the decay of $^{127}$Cs, and not from $^{127}$Ba. This way the problems caused by the relatively short half-life of $^{127}$Ba can be avoided. The cross section determined this way is of course the sum of the \ran\ and \rap\ cross sections. However, as stated above, the \rap\ gives only a very small contribution to the measured yield. This was taken into account in the determination of the \ran\ cross section and its uncertainty (see Sec.\,\ref{sec:results}).

\section{Experimental procedure}
\label{sec:exp}

\subsection{Xe gas target}
\label{sec:target}

\begin{figure}
  \begin{center}
  \epsfig{bbllx=5,bblly=541,bburx=379,bbury=755,file=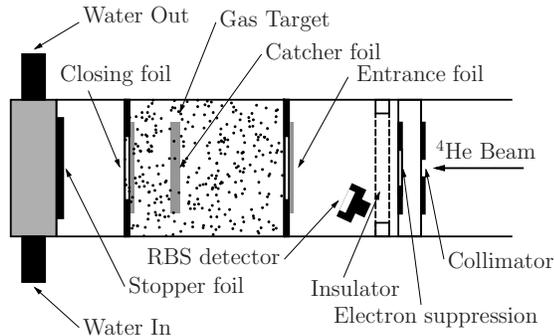,width=8cm}
  \caption{Schematic view of the gas target system. In comparison to Fig.\ 1 of \cite{bor12}, a new vacuum-chamber section was introduced to avoid total beam-energy deposition in the gas target. (Not drawn to scale.)}
  \label{Target}
  \end{center}
\end{figure}

Since xenon is a noble gas, a gas target must be used for cross section measurements of $^{124}$Xe. Recently, a thin window gas cell for activation cross section measurements was developed at Atomki. The same initial configuration was already used for the measurement of the $^{3}$He($\alpha$,$\gamma$)$^{7}$Be cross section \cite{bor13}. Details of this gas target system were published elsewhere \cite{bor12}. Here only the most important modifications relevant for the present experiment are summarized and the scheme of the system is shown in Fig.\ \ref{Target}.

\begin{figure}
  \begin{center}
  \epsfig{bbllx=65,bblly=306,bburx=532,bbury=540,file=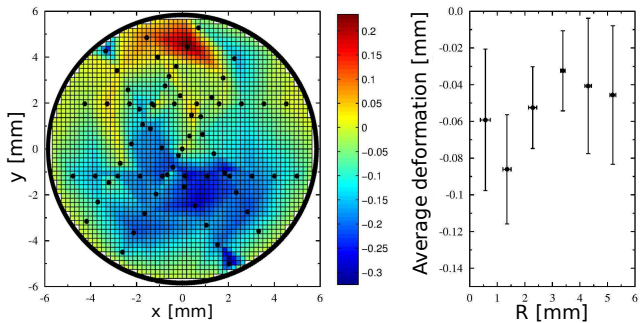,width=8.6cm}
  \caption{Left side: Deformation of the 6.6$\mu$m Al foil measured off-site at 10\,mbar pressure. Black dots show the measured positions. Negative values (colored with blue) mean increase of the target length. The positive values (colored with red) are consequences of a surface contamination producing reflection. Right side: Average deformation of the foil as the function of the distance R from the center of the foil. (Color online)}
  \label{deform}
  \end{center}
\end{figure}

The enriched xenon gas was filled into a gas cell closed on both sides by 6.6\,$\mu$m thick Al foils. The precise thickness of the Al window was measured with $\alpha$-energy loss to a precision of 5\,\%. The closing foil separates the vacuum chamber of the water-cooled beam stop and the gas cell (see Fig.\,1). This way only 10\% of the total beam energy deposits inside the cell. The total length of the cell was about 20\,mm and inside, at a distance of 10\,mm from the entrance window, a 6.6\,$\mu$m Al foil was placed. This replaceable foil served as a catcher for collecting the produced radioactive reaction products. The cell was filled to a typical pressure of 5\,mbar $^{124}$Xe measured by a MKS 09XX11 Baratron-type manometer. After the filling the cell was closed and no further xenon gas was added during an irradiation period. The pressure in the cell was monitored continuously and a steady pressure increase of typically 0.5\,mbar/h was observed caused by the penetration through the sealing and/or desorption from the stainless steel wall into the cell. The pressure was never allowed to exceed 8 mbar. The temperature of the gas was kept constant and defined by the cooling water temperature flowing through the body of the cell (see Fig.\,1 of Ref.\,\cite{bor12}).
The pressure inside the gas cell caused an elastic deformation of the Al foil entrance window which increased the cell length and thereby the number of target nuclei at a given pressure. This deformation was measured off-site with a laser distance sensor (OMRON ZX2-LDA100) and was found to increase the cell length by 0.2\,mm in the worst case at 10\,mbar pressure. From Fig.\ \ref{deform} it is obvious that at this low pressure the foil deformation was mainly caused by the fastening method of the foil and not the pressure difference. The right side of Fig.\ \ref{deform} shows the typical off-site measured deformation values as the function of the distance $R$ from the center of the foil at 10\,mbar pressure. During the  irradiation the deposited  energy can also cause foil deformation. It was found in \cite{deform}, however, to be negligible compared to the measured one.

The isotopic abundance of $^{124}$Xe in natural Xenon is very low, only 0.0952\,\%. Therefore, highly enriched $^{124}$Xe gas with 99.97\,\% enrichment was used (provided by ISOFLEX, USA, certificate No. 54-01-124-4168). The high enrichment means that radioactive species produced on heavier xenon isotopes by the $\alpha$ bombardment could not be observed. 

\subsection{Irradiations}
\label{sec:irrad}

The $\alpha$ beam for the irradiations was provided by the MGC-20 cyclotron of Atomki \cite{cyclotron}. The investigated energy range between E$_\alpha$\,=\,11 and 15\,MeV was covered with 0.5\,MeV steps. The uncertainty of the beam energy after the cyclotron switching magnet is about 0.3\%. The typical beam intensity was kept below 0.5\,p$\mu$A to avoid damage of the foils. The whole target chamber including the gas cell was insulated from the rest of the beam line and served as a Faraday cup to measure the collected charge with charge integration. In order to take into account in the activation analysis the fluctuation of the beam intensity, the collected charge was recorded in multichannel scaling mode with one minute time base. The typical lengths of the irradiations were between 3 (E$_\alpha$\,=15\,MeV) and 18 (E$_\alpha$\,=11\,MeV) hours, with 9e-3\,C and 4e-2\,C cumulative charge, respectively.

An ion-implanted Si particle detector was built into the chamber in front of the gas cell at a backward angle of 165$^\circ$. This detectors was used to detect the $\alpha$ particles backscattered on the gas cell entrance window and on the xenon gas (see Figs.\,1 and 5b of Ref.\,\cite{bor12}). The continuous monitoring of the xenon backscattering peak ensured that the xenon content of the gas cell was constant in time despite the increasing pressure mentioned above.

GEANT 4 \cite{Geant4} simulations were carried out for the motion of the beam particles and the backscattered reaction products in the gas cell. It was found that 95\,\% of both the \xeag\ and \xean\ reaction products reach the catcher foil within a circular area of 20\,mm in diameter and have at least 10\,keV kinetic energy in the case of the lowest bombarding energy. The total amount of activated nuclei has been corrected with the results of the simulations and the uncertainities were taken into account in the analysis.

\subsection{Detection of the induced $\gamma$-activity}
\label{sec:detection}

After the irradiations the catcher foil was removed from the gas cell and transported to the off-line counting setup where the $\gamma$ radiation following the $\beta$ decay of the reaction products was measured. The $\gamma$ measurements started typically one hour after the end of the irradiation. After one hour the $^{127}$Ba decayed almost completely to $^{127}$Cs (see Table I) and also some short-lived, disturbing activities vanished. 

A 100\,\% relative efficiency HPGe detector (Det 1) and a LEPS -- Low Energy Photon Spectrometer -- detector (Det 2) were used for the $\gamma$ measurements in complete 4$\pi$ low-background shieldings. Below 11.5\,MeV bombarding energy -- because of the inadequate background condition of Det 1 -- Det\,2 was used.   At 12\,MeV $\alpha$ energy the detectors were cross-checked. (See Fig.\,4.) The absolute efficiency of the detectors were measured with several calibrated radioactive sources which were used also for the energy calibration. In order to be able to measure cross sections as small as possible, the measurements were carried out at close geometry, with the catcher foils being placed at 1\,cm distance from the end cap of the detector. Although the true coincidence summing effect is less than 10\,\% in the case of $^{127}$Cs and $^{127}$Ba, the relative close and far geometry method, as described in Ref.\,\cite{hal12}, was used to determine the precise efficiency free from the coincidence summing effect.

\begin{figure}
  \begin{center}
  \epsfig{bbllx=23,bblly=523,bburx=342,bbury=753,file=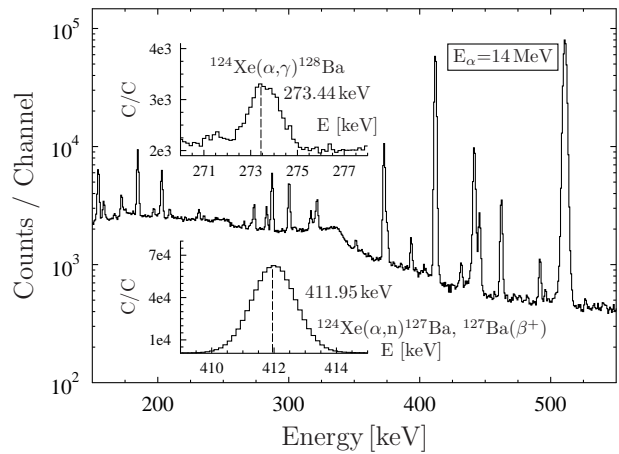,width=8cm}
  \caption{Typical $\gamma$ spectrum measured with Det\,1 on a catcher irradiated with an $\alpha$ beam of 14.0\,MeV. The two $\gamma$ peaks used for the analysis are indicated. The other peaks can be identified as laboratory background or as the effect of the activated impurities in the catcher foil \citep{AlFoil}.}
  \label{fig:spectrum}
  \end{center}
\end{figure}

The length of the $\gamma$ countings varied between several hours and several days. The spectra were stored regularly in order to follow the decay of the reaction products. Fig.\,\ref{fig:spectrum} shows a typical $\gamma$ spectrum taken on a catcher irradiated at E$_\alpha$=14\,MeV. The two $\gamma$ lines used for the analysis are indicated.

\section{Results and discussion}
\label{sec:results}

\subsection{Experimental reaction cross sections}

Table\,\ref{tab:results} shows the measured cross sections of the two studied reactions \xeag\ and \xean. The effective center-of-mass energies (second column) were calculated from the primary beam energy by taking into account the energy loss of the beam in the Al entrance foil and in the target gas itself. The effective energy in the gas was calculated using the method described in Ref.\,\cite{rolfs}. However, since the energy loss of the $\alpha$ beam is relatively small in the gas (about 6\,keV at 11\,MeV bombarding energy in the 10\,mm long gas target), the cross section does not change much in this energy range and hence the effective energy corresponds with good approximation to the middle of the target. The uncertainty of the effective energy contains the uncertainties of the primary beam energy, the Al entrance foil thickness, the $\alpha$ stopping power in Al and Xe, and the energy straggling of the beam when passing through the foil and gas.

\begin{table}
\centering
\caption{\label{tab:results} Measured cross sections of the \xeag\ and \xean\ reactions. At 12\,MeV energy only data from Det\,1 was used.}
\begin{tabular}{crclrclrcl}
\hline
E$_\alpha$ & \multicolumn{3}{c}{E$^{\mbox{\small eff}}_{\mbox{\small cm}}$} & \multicolumn{6}{c}{cross section [$\mu$barn]}\\
\,[MeV] & \multicolumn{3}{c}{[MeV]} &  \multicolumn{3}{c}{\xeag}  & \multicolumn{3}{c}{\xean}\\
\hline
15.0	&	14.0	&$\pm$&	0.1	&	785	&$\pm$&	143	&	12032	&$\pm$	&	1585\\
14.5	&	13.6	&$\pm$&	0.1	&	456	&$\pm$&	82	&	4065	&$\pm$	&	540\\
14.0	&	13.1	&$\pm$&	0.1	&	283	&$\pm$&	41	&	2051	&$\pm$	&	275\\
13.5	&	12.6	&$\pm$&	0.1	&	117	&$\pm$&	24	&	684	&$\pm$	&	92\\
13.0	&	12.1	&$\pm$&	0.1	&	84	&$\pm$&	17	&	305	&$\pm$	&	41\\
12.5	&	11.6	&$\pm$&	0.1	&	43	&$\pm$&	10	&	-	&-	&	-	\\
12.0	&	11.1	&$\pm$&	0.1	&	28	&$\pm$&	4	&	-	&-	&	-	\\
11.5	&	10.6	&$\pm$&	0.1	&	15	&$\pm$&	3	&	-	&-	&	-	\\
11.0	&	10.1	&$\pm$&	0.1	&	5.3	&$\pm$&	0.6	&	-	&-	&	-	\\
\hline
\end{tabular}
\end{table}

The uncertainty of the cross section is calculated by taking into account the following partial errors: number of projectiles (3\,\%, from the precision of the charge measurement), number of target atoms (2\,\%, from the uncertainty of pressure and temperature measurement), the change of the physical dimensions of the gas cell (0.4\,\%, from the deformation measurements), $\gamma$-detection efficiency (7-11\,\%), counting statistics ($<8$\,\% for the ($\alpha$,$\gamma$) and $<1$\,\% for the ($\alpha,n$) reaction) and the ($\alpha$,$p$) reaction's contribution ($<5$\,\%)).

The lowest energy irradiation (E$_{cm}^{\mbox{eff}}$\,=\,10.1\,MeV) was below the \ran\ reaction threshold (10.83\,MeV), taking into account the energy loss in the entrance foil. In the case of the next three energies (E$_\alpha$\,=\,11.5, 12 and 12.5\,MeV) the yield of the \ran\ reaction channel was too low. Therefore, the \xean\ cross section was determined only above these energies while the \xeag\ cross section was successfully measured in the whole studied energy range.

\begin{figure}
  \begin{center}
  \epsfig{bbllx=23,bblly=523,bburx=342,bbury=753,file=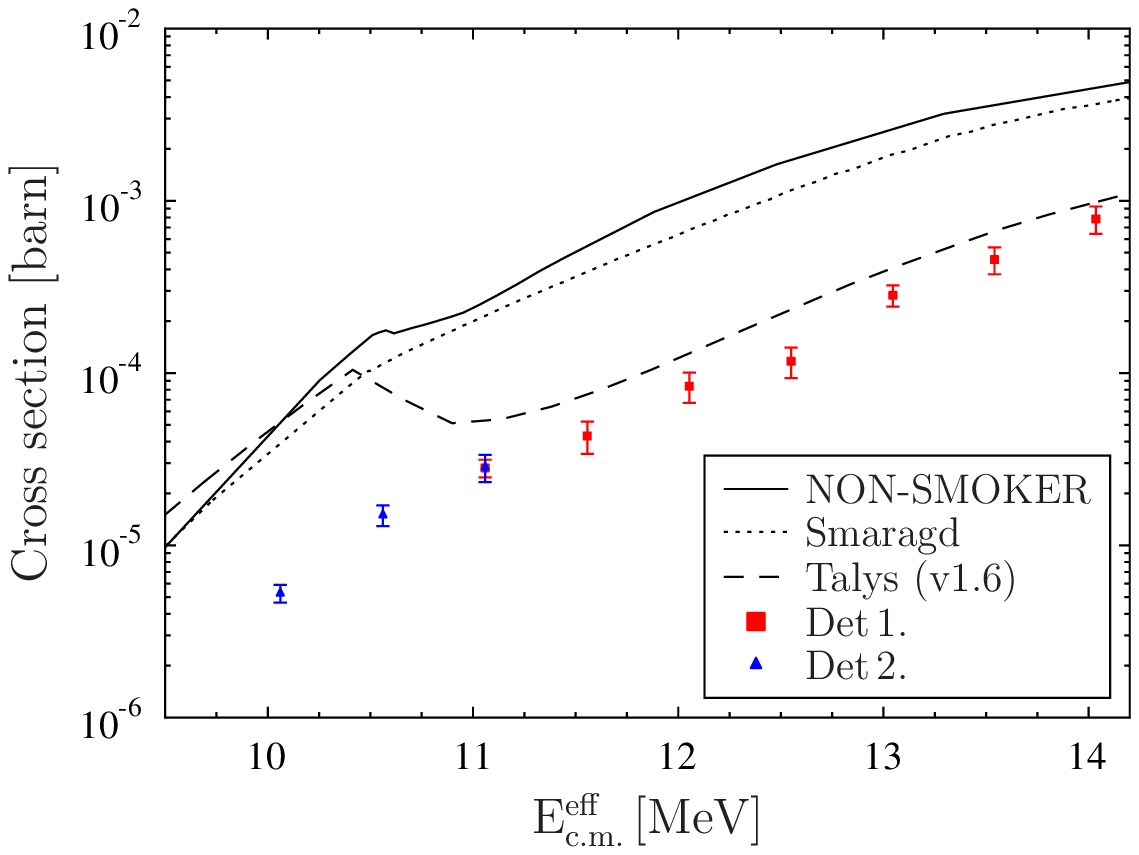,width=8cm}
  \caption{(Color online) Reaction cross section of the \xeag\ reaction.}
  \label{fig:agres}
  \end{center}
\end{figure}

\begin{figure}
  \begin{center}
  \epsfig{bbllx=23,bblly=523,bburx=342,bbury=753,file=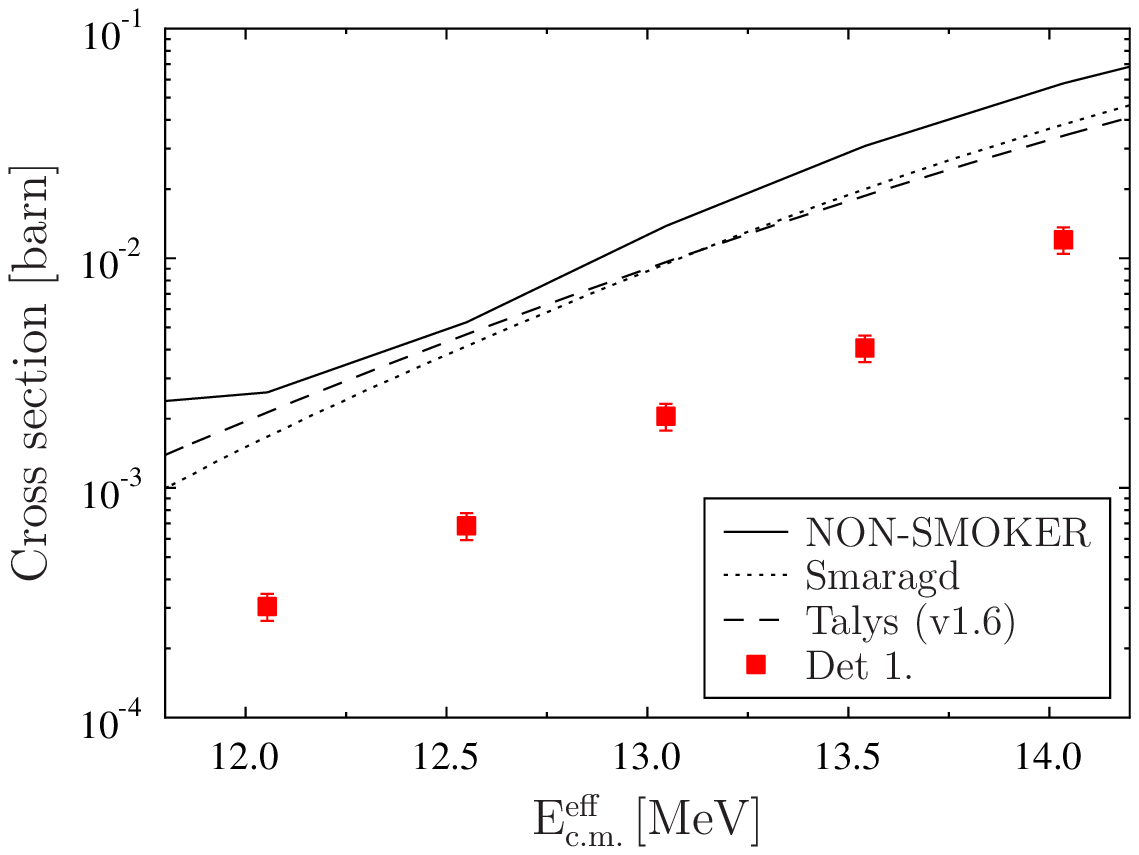,width=8cm}
  \caption{(Color online) Reaction cross section of the \xean\ reaction.}
  \label{fig:anres}
  \end{center}
\end{figure}

Figures\,\ref{fig:agres} and \ref{fig:anres} show the measured cross section of the \xeag\ and \xean\ reactions, respectively, as astrophysical $S$ factor \cite{rolfs}. The results of statistical model calculations for the cross sections obtained with the default settings of the Hauser-Feshbach codes NON-SMOKER \cite{adndt00,adndt01}, TALYS \cite{TALYS}, and SMARAGD (version 0.9.3s) \cite{SMARAGD} are also included in the figures. Apart from differences in the internal numerics, the codes mainly differ in the input used to determine the transmission coefficients which are the central quantities in the Hauser-Fesbach approach \cite{haufesh,raureview}. Of special relevance here is the optical $\alpha$+nucleus potential, as discussed in more detail below. Currently, the rates of \cite{adndt00} are used in most astrophysical studies. 

As can be seen in Figs.\ \ref{fig:agres} and \ref{fig:anres}, the shown NON-SMOKER and SMARAGD cross sections reproduce the measured energy dependence well but are larger by about a factor of 5 in both reactions. The TALYS calculation using the standard input parameters of the code version 1.6 gives similar ($\alpha$,n) cross sections as NON-SMOKER and SMARAGD. The unusual energy dependence of the TALYS ($\alpha$,$\gamma$) cross section, on the other hand, is not reproduced by the measured data.

\subsection{Comparison to theory and implications for the stellar rate}

In this section we want to provide a more detailed analysis aimed at understanding the discrepancy between theory and experiment, and its impact on the stellar rates.

\begin{figure}
\includegraphics[angle=-90,width=\columnwidth]{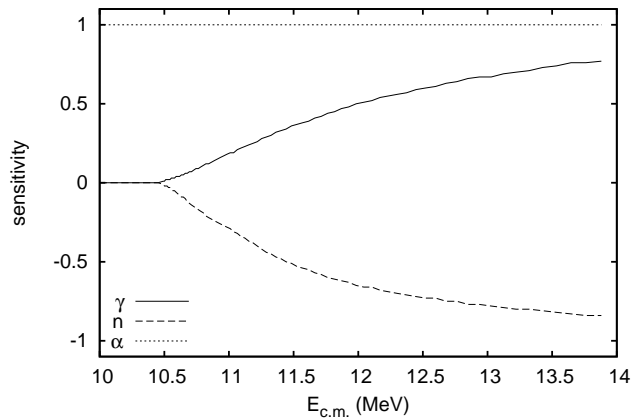}
\caption{\label{fig:sensiag}Sensitivity of the \xeag\ cross section to variations of the neutron-, $\alpha$-, and $\gamma$ width as function of c.m.\ energy.}
\end{figure}

\begin{figure}
\includegraphics[angle=-90,width=\columnwidth]{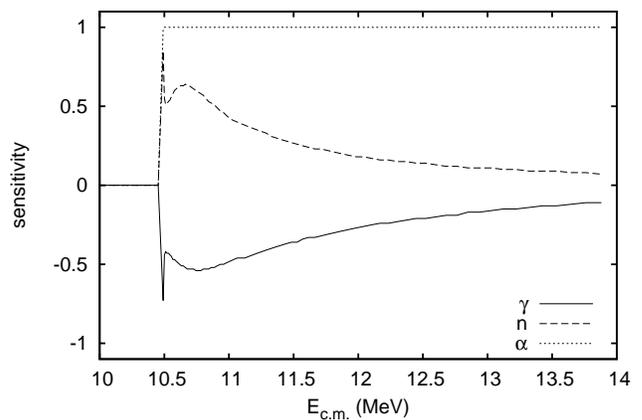}
\caption{\label{fig:sensian}Sensitivity of the \xean\ cross section to variations of the neutron-, $\alpha$-, and $\gamma$ width as function of c.m.\ energy.}
\end{figure}

To understand the source of the differences, and whether it is important for the astrophysical rate, it is necessary to consider the sensitivity of the calculated cross sections on the various nuclear properties entering the reaction model. Central to the Hauser-Feshbach model are the averaged widths, computed from transmission coefficients which, in turn, are obtained from solutions of the Schr\"odinger equation using effective interaction potentials or, for photon transmission coefficients, from the knowledge of photon strength functions and nuclear level densities \cite{raureview}. For reactions in astrophysics, mostly averaged neutron-, proton-, $\alpha$-, and $\gamma$ widths have to be considered. For the reactions considered here, the impact of the proton width is negligible. Figures \ref{fig:sensiag} and \ref{fig:sensian} show the sensitivities of the \rag\ and \ran\ cross sections, respectively, to variations of the remaining widths. The sensitivity is defined as in \cite{rau12} and can vary between 0 (no sensitivity) and 1 (full sensitivity), i.e., the cross section $\sigma$ will be changed by a factor $s\,f$, where $s$ is the sensitivity and $f$ the factor with which the width was varied. For $s<0$, the varied cross section is given by $\sigma/(s\,f)$.

The sensitivities shown here differ from those given in \cite{rau12} because it turned out that the excited states given in \cite{NuDat}, which were used for the calculations in \cite{rau12}, do not define a complete set of states in $^{127}$Ba. Due to the endothermic nature of \xean\, this strongly affects the resulting cross sections in the measured energy range. (Similar problems due to incomplete sets of experimentally determined excited states were also discussed in \cite{gur15,kiss12}.) In the present work, the incomplete set of excited states was neglected in the SMARAGD calculations and instead the nuclear level density of \cite{rtk} with the parity distribution of \cite{darko07} was used above the ground state.

As can be seen in Figs.\ \ref{fig:sensiag} and \ref{fig:sensian}, the cross sections of both reactions are equally sensitive to the $\alpha$ width across the measured energy range. On the other hand, the sensitivities to variations of the neutron- and $\gamma$ widths are energy dependent and width variations also act oppositely in the two reactions. It should be noted that the three different calculations shown in Figs.\ \ref{fig:agres} and \ref{fig:anres} use different descriptions of nuclear properties.

Since the SMARAGD cross sections are above the experimentally determined ones in both reactions, a consistent rescaling can only be achieved by altering the $\alpha$ width. Even more constraining is the fact that it was possible to obtain \rag\ data below the \ran\ threshold, where the neutron width does not contribute and the $\gamma$ width is larger than the $\alpha$ width, leading to a dependency of the cross section solely on the $\alpha$ width. Using the standard $\alpha$+nucleus optical potential by \cite{McF66}, which was also used in the NON-SMOKER rates \cite{adndt00}, the SMARAGD cross sections shown in Fig.\ \ref{fig:sensiag} were obtained, requiring a renormalization of the calculated $\alpha$ width by a factor of 0.125 to be in agreement with the experimental data point at the lowest energy.\newline
Combining this with the \ran\ data at the upper end of the investigated energy range, which again almost exclusively depends on the $\alpha$ width, allows to check whether the required scaling would be energy dependent. At the data point with highest energy, the $\alpha$ width had to be divided by three to reproduce the experimental cross section. It has to be cautioned, however, that the cross sections at this energy still are weakly dependent on the neutron- and $\gamma$ widths.
Such a strong energy dependence has not been observed yet in other \rag\ reactions studied in connection with the $\gamma$ process. It cannot, for example, be reproduced by the energy-dependent optical potential introduced in \cite{sauer} and subsequently used successfully to describe further reactions (see the experimental references given in Sec.\ \ref{sec:intro}).

%
%

\begin{figure}
\includegraphics[angle=-90,width=\columnwidth]{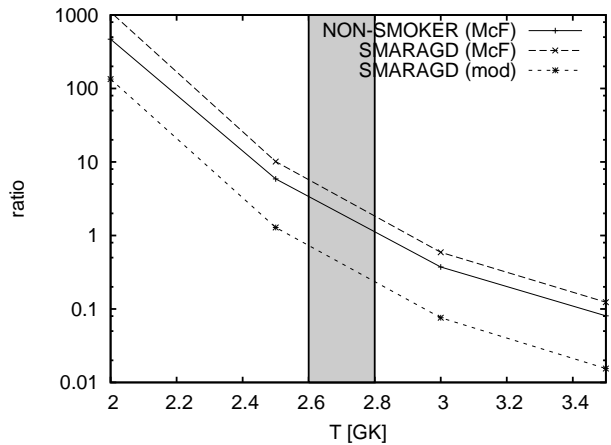}
\caption{\label{fig:branching}Ratio of the $^{128}$Ba\rga\ rate to the largest competing rate, which is \rgn\ except for \rgp\ at the lowest temperature, as function of the stellar plasma temperature $T$. The temperature range affecting the Xe-Ba region is shown by the grey box.}
\end{figure}

Further experimental data below the \ran\ threshold would be needed to constrain fully the energy dependence and to determine the $\alpha$ widths at astrophysically relevant energies due to the appearance of $\alpha$ transitions to excited states in \textit{stellar} rates. For the reaction in question, the relevant energy window lies below the measured energy range. Even without this additional information, however, it is possible to discuss the astrophysical impact of a reduced $\alpha$ width. Required for astrophysical reaction network calculations are the \textit{stellar} rates (including reactions on thermally excited nuclei) for \xeag\ and its reverse reaction which can be derived applying detailed balance \cite{raureview}. Both rates, for forward and reverse reaction, scale by the same factors and thus scaling the stellar \rag\ rate will scale the stellar \rga\ rate similarily. (Note that this only applies to \textit{stellar} rates including the effect of thermally excited nuclei, but not to g.s.\ rates and cross sections.) In the $\gamma$ process relevant for the production of $p$ nuclei, the \rga\ reaction will dominate over the capture reaction, as there are no free $\alpha$ particles to be captured. As shown in Fig.\ \ref{fig:branching} and discussed in \cite{rau06}, the stellar rate for $^{128}$Ba($\gamma$,$\alpha$)$^{124}$Xe is close to the $^{128}$Ba($\gamma$,n) and $^{128}$Ba($\gamma$,p) rates and thus may provide a deflection in the path of \rgn\ photodisintegrations. An increase in the rate would not only increase the $^{124}$Xe production through this path but also increase the sensitivity of the final abundance to this reaction. This is why a stronger dependence on this rate was found in \cite{rap06} when increasing it strongly. Our experimental data, however, rule out an increase and rather suggest a considerable suppression of this reaction path. A rate reduction by a factor of 0.125, as found with the lowest data point, already renders the \rga\ rate unimportant compared to \rgn\ in the relevant temperature range. A further reduction, although to be expected from the energy dependence extracted from the experimental data, will not change this conclusion and will have no further impact on the final abundance of $p$ nuclei. This is also the reason why Ref.\ \cite{rap06} found only a modest impact of the rate even when reducing it by a factor of 0.1.

\begin{figure}
\includegraphics[angle=-90,width=\columnwidth]{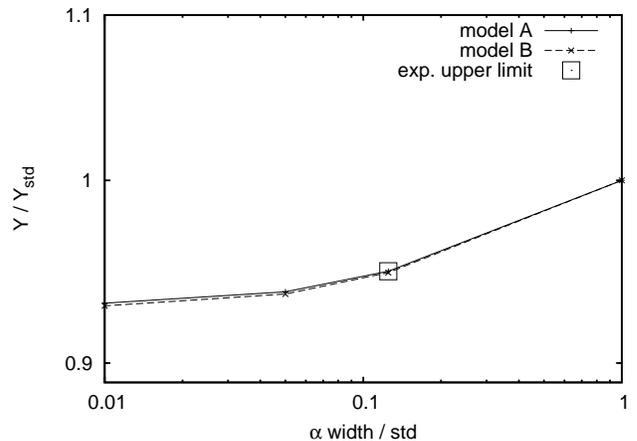}
\caption{\label{fig:stellarmodels}Abundance $Y$ of $^{124}$Xe obtained with rates based on various $\alpha$ widths as function of $\alpha$ width variation factor. A similar dependence was obtained in two stellar models A and B (see text for details). The value inferred from the experimental data as an upper limit of the $\alpha$ width is marked by the box. Note that the ordinate is given on a logarithmic scale.}
\end{figure}

To confirm this, we have performed full $\gamma$-process network calculations with different \xeag\ rates, using two approaches to simulate nucleosynthesis in the explosion of a 25 $M_\odot$ star. The default NON-SMOKER rate (standard rate) was taken from \cite{adndt00}, as it was also used in \cite{rap06} and in rate set A of \cite{rau06}. As a second trial rate, we use the rate obtained with the SMARAGD code when using the potential by \cite{McF66} but renormalizing the obtained $\alpha$ width by a factor of 0.125 as suggested by the experimental data. A third and fourth rate were obtained by using factors of 0.05 and 0.01, respectively. In terms of astrophysical models, the mass zones and trajectories from the stellar model of \cite{rayet95,hashi} were employed (model A). The same model was applied in \cite{rap06}. Recently it was pointed out, however, that the zoning grid in this model is too coarse to follow the production of $p$ nuclei in detail \cite{raunobhir}. Therefore we studied the $\gamma$ process also in a second model with a finer grid of mass shells (model B) \cite{rau02,raunobhir}. The $^{124}$Xe abundance changes for the eight calculations (four rates each in two models) are shown in Fig.\ \ref{fig:stellarmodels}. Both stellar models show the same dependence of the final $^{124}$Xe abundance on the $\alpha$ width. As expected from the inspection of the rate ratios, the production decreases when using the modified rates. The reduction is very moderate, however, and much smaller than the ratios of the rates with respect to the standard rate. This is in line with what was found in Fig.\ \ref{fig:branching}. Any further reduction also would not alter the production factor anymore because the \rga\ rate is already smaller than the \rgn\ one and thus its absolute value is not significant anymore.

\section{Conclusion and summary}
\label{sec:conclusions}

The reaction cross sections of \xeag\ and \xean\ were measured close to the \ran\ threshold using the activation method. It was found that the experimental cross sections for both reactions are smaller than the ones predicted by the reaction model which has been used in previous $\gamma$-process studies. This is in line with the findings of previous \rag\ and \ran\ experiments on other $p$ nuclei, which also showed smaller cross sections than predicted (see \cite{net13, kis14, yal15} and references in \cite{ERCRep}).

Although the covered energy range is slightly above the astrophysically relevant energies, it is possible to evaluate the astrophysical impact of the measurement. The \rag\ cross section below the threshold can be taken as an upper limit for the calculation of the astrophysically relevant $^{128}$Ba\rga\ reaction rate. Since the new rate is already below the $^{128}$Ba\rgn\ rate, it was possible to show that this reaction path cannot contribute significantly to the synthesis of the $p$ nucleus $^{124}$Xe.

A further reduction of the $^{128}$Ba\rga\ rate does not change this conclusion. It can be affected, however, by modified $^{128}$Ba\rgn\ and $^{128}$Ba\rgp\ rates. The values of these rates used in current astrophysical simulations are still based on theoretical predictions by \cite{adndt00,SMARAGD}. An experimental investigation of the respective \rng\ and \rpg\ cross sections close to the astrophysically relevant energy window would be desirable. Unfortunately, this proves to be very difficult because both $^{127}$Ba and $^{127}$Cs are unstable and any measurements will have to await further technical development.

\begin{acknowledgments}
This work was supported by OTKA (K101328, K108459). G.G. Kiss acknowledged support from the J\'anos Bolyai Research Scholarship of the Hungarian Academy of Sciences. T.\,Rauscher was supported by the UK STFC grant ST/M000958/1 and the ERC Advanced grant GA 321263-FISH. J.\,Gorres and M.\, Wiescher were supported by the National Science Foundation through grant No. PHY 1068192. This work was supported by The Joint Institute for Nuclear Astrophysics through grant No. PHY 0822648.
\end{acknowledgments}

\end{document}